\documentclass[aps,twocolumn,tightenlines]{revtex4}
\usepackage{graphicx}
\begin{document}
\title{Pattern formation in mixtures of ultracold atoms in optical lattices}
\author{M.~M.~Ma\'ska}
\affiliation{Department of Theoretical Physics, Institute of Physics, University of Silesia, 40-007 Katowice, Poland}  \author{R.~Lema\'nski}
\affiliation{Institute of Low Temperature and Structure Research, Polish Academy of Science, 50-422 Wroc{\l}aw, Poland}
\author{J.~K.~Freericks}
\affiliation{Department of Physics, Georgetown University, Washington, DC 20057, U.~S.~A.}
\author{C.~J.~Williams}
\affiliation{Joint Quantum Institute, National Institute of Standards and Technology, Gaithersburg, MD 20899-8420}

\date{\today}

\begin{abstract}
Regular pattern formation is ubiquitous in nature; it occurs in biological, physical, and materials science systems.  
Here we propose a set of experiments with ultracold atoms that show how to examine different types of pattern formation.  In particular, we show how one can see the analog of labyrinthine patterns (so-called quantum emulsions) in mixtures of light and heavy atoms (that tend to phase separate) by tuning the trap potential 
and we show how complex geometrically ordered patterns emerge (when the mixtures do not phase separate), which could be employed for low-temperature thermometry.  The complex physical mechanisms for the pattern formation at zero temperature are understood within a theoretical analysis 
called the local density approximation.
\end{abstract}


\maketitle

We choose to examine a physical realization of mixtures of ultracold atoms that illustrate pattern formation---a mixture of fully polarized light fermionic atoms with heavy atoms that are either fully polarized fermions or (nearly) hard core bosons.  This choice implies that there are only two possibilities for the occupation of each atomic species on an optical lattice site: either no particle or one particle of each species is allowed.  We also assume that the heavy particles are essentially localized, but the theoretical system nevertheless has the freedom to rearrange itself amongst the different possible configurations of the heavy particles (which is analogous to the Ising model of magnetism, where the system does not allow \textit{dynamic} quantum-mechanical spin flips but different spin configurations are sampled in a statistical-mechanical sense via an averaging over all possible states). Most importantly, there is an interspecies interaction when both atoms lie on the same optical lattice site; this interaction is ultimately responsible for the pattern formation that emerges in these systems (pattern formation will occur for either attractive or repulsive interspecies interaction which are related via a particle-hole transformation in the homogeneous phase diagram; we focus on repulsive interactions here---note that attractive interactions may have technological use in creating polar molecules on a lattice). 
Experimental realizations of such a system can be found, for example, in mixtures of heavy fermionic atoms like  $^{87}$Sr, $^{171}$Yb, or $^{173}$Yb with light fermionic atoms like $^6$Li or $^{40}$K.

The light atoms can hop between nearest neighbors on the optical lattice, and they feel an inhomogeneous trapping potential (assumed to be harmonic here) that keeps them close to the center of the lattice. These atoms will be described by spinless fermionic creation $c^\dagger_i$ and annihilation $c^{}_i$ operators associated with each lattice site $i$ (because they are fully polarized). The heavy atoms do not hop, so they are described by classical occupation numbers $w_i$ which equal zero or one at each lattice site. The heavy particles are in an inhomogeneous trapping potential (also assumed to be harmonic, but with an independent trap frequency).  Finally, there is an on-site interaction between the two atoms denoted by $U$.  The Hamiltonian then becomes
\begin{eqnarray}
 \mathcal{H}&=&-J\sum_{\langle i,j\rangle}(c^\dagger_ic^{}_j+c^\dagger_jc^{}_i)
+\sum_i(V_i-\mu)c^\dagger_ic^{}_i\nonumber\\
&+&\sum_i(V_i^h-\mu^h)w_i
+U\sum_ic^\dagger_ic^{}_iw_i,
\label{eq; hamiltonian}
\end{eqnarray}
which is called the spinless Falicov-Kimball model~\cite{falicov_kimball} and has been proposed within the ultracold atom context by Ates and Ziegler~\cite{ziegler}. The symbol $J$ is the nearest-neighbor hopping integral, and the angle brackets denote a summation over nearest neighbor pairs, while the symbols $V_i$ ($V_i^h$) are the harmonic trapping potentials for the light (heavy) atoms, and $\mu$ ($\mu^h$) are the corresponding chemical potentials.  
For concreteness, we work on a two-dimensional square lattice, with lattice spacing $a$.

The spinless Falicov-Kimball model is known to have a rich phase diagram in the homogeneous case (when the trapping potentials are set equal to zero).  On a square lattice, when the light and the heavy atoms each occupy one half of the lattice sites, the system is known to form a checkerboard pattern that alternates between the light and the heavy atoms for all values of $U$~\cite{lieb_kennedy,brandt-schmidt}.  When the interaction is large and repulsive $U\rightarrow+\infty$, the system phase separates (all light atoms move to one side of the lattice, and the heavies move to the other side) whenever the total number of atoms is less than the number of lattice sites~\cite{freericks_lieb_ueltschi}.  Elsewhere, the system displays a complex phase diagram~\cite{watson_lemanski}, which includes stripe-like phases and more complex two-dimensional patterns~\cite{lemanski_freericks_banach}.   We expect the inhomogeneous system in a trap will display
different phases at different spatial locations as the local chemical potential sweeps through the phase diagram.  
This should produce a ``generalized wedding-cake-like'' structure~\cite{zoller,batrouni}, but with the different ``layers'' corresponding to different complex ordered patterns.

We use a separate harmonic trap for each of the two atomic species (whose minima are both located at the origin in space), and treat the
trap frequency as a tunable parameter; this can be realized experimentally by using optical generated traps.  By making one trap sharper, we 
squeeze the particles to lie in the center, and we can tune through
a number of different stable phases.  The temperature at which 
the ordering occurs also
changes with the changing patterns, so it is possible that these studies 
could lead to
interesting ways to perform thermometry.  The phases that appear also depend
on the ratio of the interparticle interaction $U$ to the intersite
hopping $J$ of the light atoms.  The ratio of these two can be tuned either
via an interspecies Feshbach resonance that can change $U$, or by tuning the magnitude of
$J$.  

In our calculations, we take a square
lattice that is a $50\times 50$ cluster occupied
with 625 light (mobile) and 625 heavy
(localized) atoms. The light atoms are confined in a trap potential of the 
following form
\begin{equation}
V_i=J\left [\frac{\hbar\omega}{2Ja}\right ]^2(x_i^2+y_i^2),
\label{eq: trap}
\end{equation}
where the trap frequency $\omega$  determines the curvature of the trap potential
[site $i$ has position $(x_i,y_i)$] and we will vary $1/30\le \hbar\omega/2J\le 1/12.9$. The heavy atoms are trapped in a similar potential with a fixed curvature corresponding to $\hbar\omega^h/2J=1/30$.

Our analysis is based on a quantum Monte Carlo (QMC) approach and on
a local density approximation~\cite{lda} (LDA) approach.  The QMC algorithm is identical to that in the homogeneous case~\cite{dutch,maska} except that the hopping matrix is supplemented with a diagonal term associated with the trapping potential.  The LDA, at $T=0$, can be constructed
from the homogeneous grand canonical ground-state phase diagram~\cite{watson_lemanski} as a function of the chemical potentials.
We determine
the local chemical potentials for each lattice site by subtracting the trap
potential at that lattice site, and then we map out the quantum-mechanical phase diagram for each spatial coordinate within the trap.  The chemical potentials are then adjusted to produce the correct total number of heavy and light atoms in the lattice.

There are a number of ways to create an experimental system that behaves like the spinless Falicov-Kimball model.  First, one can slowly adjust the depth of the optical lattice for the heavy atoms, to allow different  configurations of the heavy atoms to be sampled as the lattice turns on, and then freeze in one of those configurations. Second, one can work with heavy atoms that are not fully localized, but do hop slow enough that it is safe to ignore the quantum-mechanical energy associated with their motion through the lattice.  Third, one can view each experimental configuration of the localized atoms as a frozen snapshot chosen according to the equilibrium probability distribution of the Falicov-Kimball model, so averaging over different experimental runs will produce the (equilibrium) results of the model. An imaging experiment can then measure the configuration of the atoms at a specific moment in time.  We simulate this type of experimental measurement by taking representative snapshots of the current QMC configuration
(after thermalization), to see what a ``typical'' density profile is for each atomic species.

We first show results of these snapshots for the case where $U=5J$ (see Fig.~\ref{fig: phase-sep}), 
which is in the moderately strong interaction regime
(the noninteracting bandwidth on a square lattice with nearest-neighbor hopping is $8J$), where the system has a significant tendency to phase separate.  We see the character of the phase separation change from
having the heavy particles at the center, to having them at the periphery
of the trap, as is expected for two species that phase separate and lie in different curvature traps.  What is more interesting are the patterns that emerge in the QMC snapshots in the crossover regime, where the transition from the heavies being at the center to being in the periphery occurs through a complicated intermediate state that displays an obvious analog to labyrinthine pattern formation in immiscible ferrofluids~\cite{ferrofluids}. Indeed, our LDA analysis shows that the ordered phases that create the labyrinthine patterns are the well-known axial stripe phases that appear over a wide range of the homogeneous phase diagram. Stripes along one axis merge with stripes oriented along a perpendicular axis to create the labyrinthine effect.  This behavior is also similar to the quantum emulsions seen in mixtures of different mass bosons in one dimension~\cite{cirac}.

Next we examine the weakly correlated regime with $U=J$ in Fig.~\ref{fig: u=1}.  
In this case,
we tend to see more different ordered phases in addition
to the phase separation.  The system also displays a strong
sensitivity of the pattern shape and physical extent on the temperature (particularly the heavy atoms), which suggests that this could be used for thermometry purposes if the size of the atomic cloud could be measured with a precision on the order of a lattice site.  One can also try to fit the tail of the light atom distribution (where there are no heavy atoms) to determine the temperature, as was recently done in experiments on polarized mixtures of spin-one-half fermions~\cite{ketterle}.

In the panels of Fig.~\ref{fig: u=1}, 
we show representative QMC snapshots again, this time for 
only two cases---$\hbar\omega/2J=1/17$ and $\hbar\omega/2J=1/12.9$.  For each case, two temperatures
are plotted.  One is relatively high as compared to the overall ordering temperature, while one is a very low temperature (essentially $T=0$; our results are virtually identical for $k_BT=0.0001J$ and $k_BT=0.001J$).
One of the first things to notice is that the temperature can be roughly found
just by examining the physical size of the heavy atomic cloud.  With the same
trap potential, the extent of the peripheral region of the heavy atom cloud depends
strongly on the temperature, but needs to be calibrated for experimental use.  
The shape of the 
density profile at the edges of the trap also display strong
sensitivity to $T$, and might be useful for thermometry as well---note the ordered phases that appear at the edges of both low temperature panels (b) and (d) are higher-period ordered phases whose appearance is sensitive to the $T$.

Finally, we compare the QMC results to the LDA at $T=0$ in Figs.~\ref{fig: lda} 
and \ref{fig: lda2}.  
We show plots of the LDA for 
the density as a function of radial distance from the center of the
trap for the QMC snapshots plotted above.  We compare the LDA directly
to the QMC average. Note how similar the LDA results
are with those of the QMC simulations, especially in the cases when we do not have labyrinthine patterns (the difference is probably a temperature effect); the deviations for small radii arise from vacancies in the given snapshots. Also note that when
there are ordered phases (and different traps), the LDA need not show
a monotonic dependence of the density on the distance from the center
of the trap (although the light atom density is always {\it piecewise} monotonic when the system is in one phase). 

The most direct way to observe the pattern formation is a tomography experiment with a resolution on the order of a single lattice spacing, which would provide an {\it in situ} means to directly observe the patterns.  Unfortunately, such resolution is not yet available.  Another approach is to examine Bragg scattering peaks.  In particular, when we have labyrinthine patterns forming, the structure factor will be peaked around $(\pi,0)$ and $(0,\pi)$, while the checkerboard phase will show peaks at $(\pi,\pi)$.  
The large growth of these peaks at low $T$ determine the relative area of the ordered phases 
and provide another means for thermometry of the system.  Finally, one can examine the density distributions in momentum space in time-of-flight imaging that correlate the shape of $n(k)$ with different spatial extents of the light atom clouds 
or employ noise correlation spectroscopy in time-of-flight imaging to examine the density-density correlation functions, whose momentum dependence will show peaks at wavevectors associated with particular ordering patterns.  The momentum distribution itself (not shown here) does not readily show which phases are present, but it can determine the overall spatial extent of the clouds of atoms.

In conclusion, we have shown that interesting density-wave patterns appear in mixtures of spin polarized fermionic atoms with different masses. The ordering phenomena ranges from geometrical phases to labyrinthine patterns reminiscent of viscous fingering in immiscible ferrofluids. The ordering phenomena is most striking when the heavy mass atom is localized on the lattice, but we expect these patterns to survive when the kinetic energy for the heavy atoms remains small (due to a Fermi pressure effect~\cite{ueltschi}).  Note that one needs to achieve low temperatures to see these patterns.  While this is not yet feasible in current experiments, new developments in cooling technology~\cite{cooling} make it seem likely that this phenomena will be able to be observed soon.

{\it Acknowledgments:}
M.~M.~M. acknowledges support from the Polish Ministry of Science and Higher Education under Grant No. 1 P03B 071 30. J.~K.~F. acknowledges support from the National Science Foundation under grant number DMR-0705266 and from the Army Research Office under award W911NF0710576 with funds from the DARPA OLE Program. This research used resources of the National Center for Computational Sciences at Oak Ridge National Laboratory which is supported by the Office of Science of the Department of Energy under Contract DE-AC05-00OR22725.

\vskip 0.1in


\begin{widetext}

\begin{figure*}[h]
\includegraphics[width=6.8in]{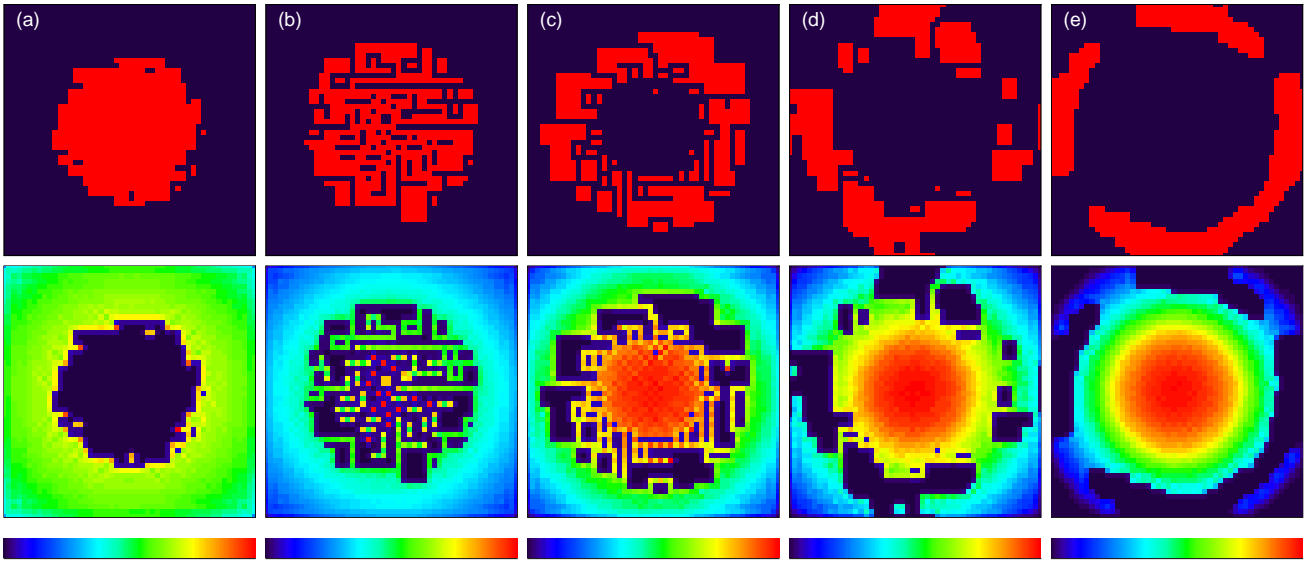}
\caption{(Color online)
Representative QMC snapshots of real-space configurations of heavy (top) and
light (bottom) atoms, respectively. All simulations have been
performed for $U=5J$ at temperature $k_BT=0.01J$. Each image pair as we move from left to right corresponds to
a different $\hbar\omega/2J$ value: (a) $1/30$; (b) $1/20$; (c) $1/18.5$; (d) $1/17$; and (e) $1/12.9$. The heavy atoms in a snapshot either occupy a site (light gray or red) or do not (black), while the light atom densities are plotted with a false-color plot. The heavies migrate from the center to the periphery as the light trap tightens (left to right).
}
\label{fig: phase-sep}
\end{figure*}

\begin{figure*}
\includegraphics[width=5.5in]{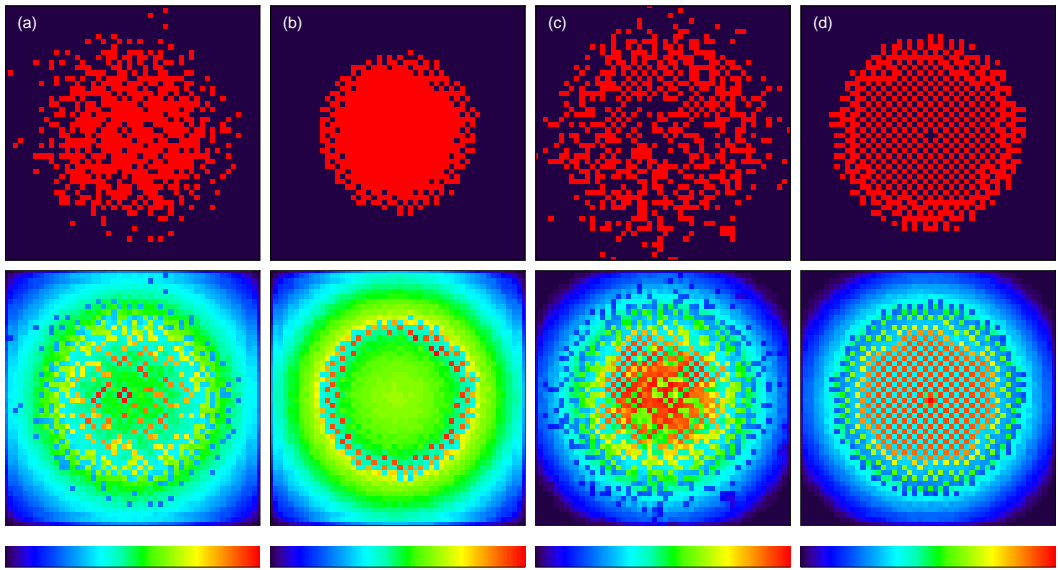}
\caption{(Color online)
Representative QMC snapshots of real-space configurations of heavy (top) and
light (bottom) atoms, respectively. These simulations were performed for $U=J$.  The parameters are as follows: (a) $k_BT=0.05J,\ \hbar\omega/2J=1/17$;
(b) $k_BT=0.0001J,\ \hbar\omega/2J=1/17$; (c) $k_BT=0.05J,\ \hbar\omega/2J=1/12.9$;
(d) $k_BT=0.0001J,\ \hbar\omega/2J=1/12.9$. 
\label{fig: u=1}
}
\end{figure*}

\begin{figure*}[h]
\includegraphics[width=5.6cm]{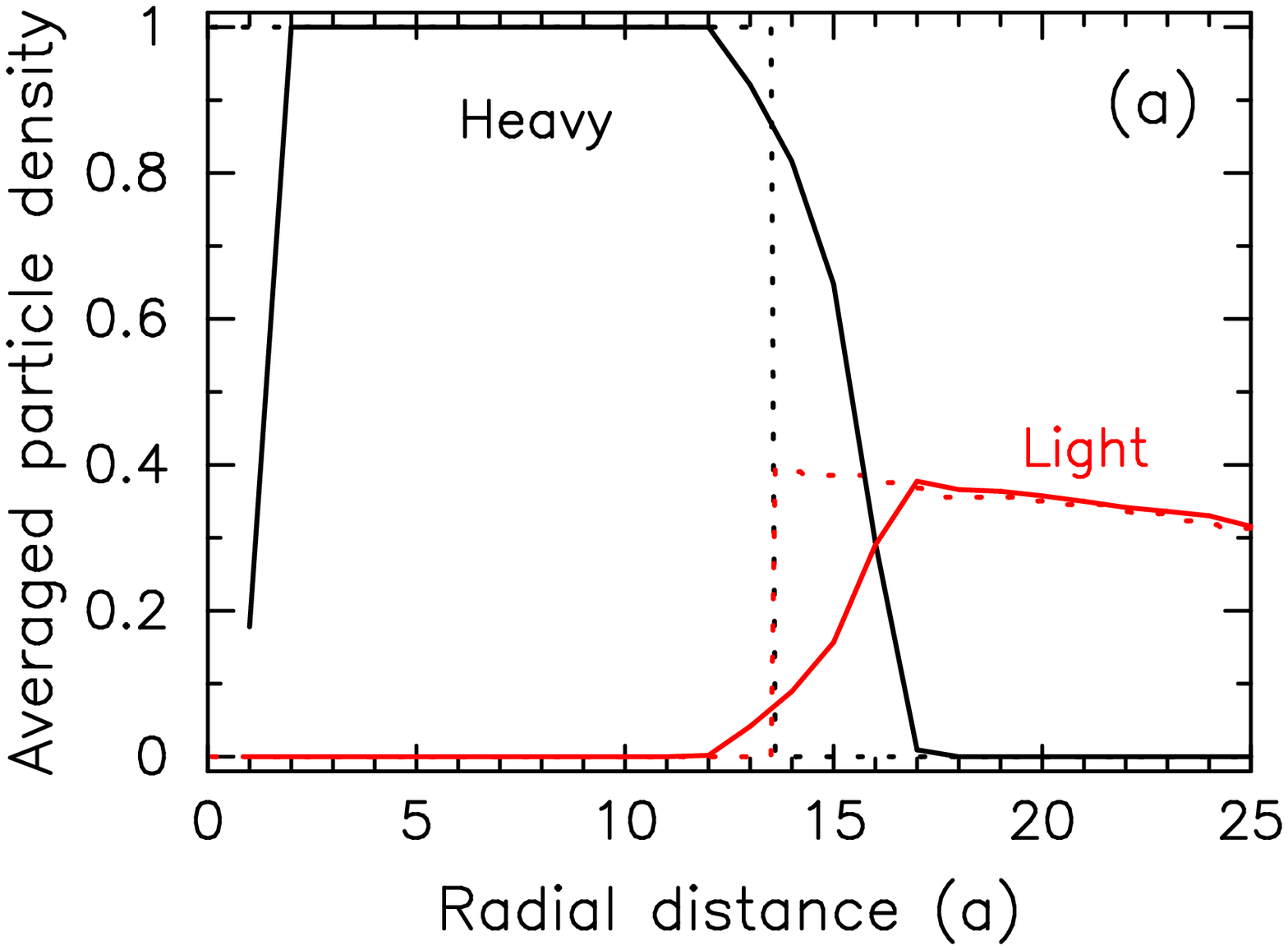}
\includegraphics[width=5.3cm]{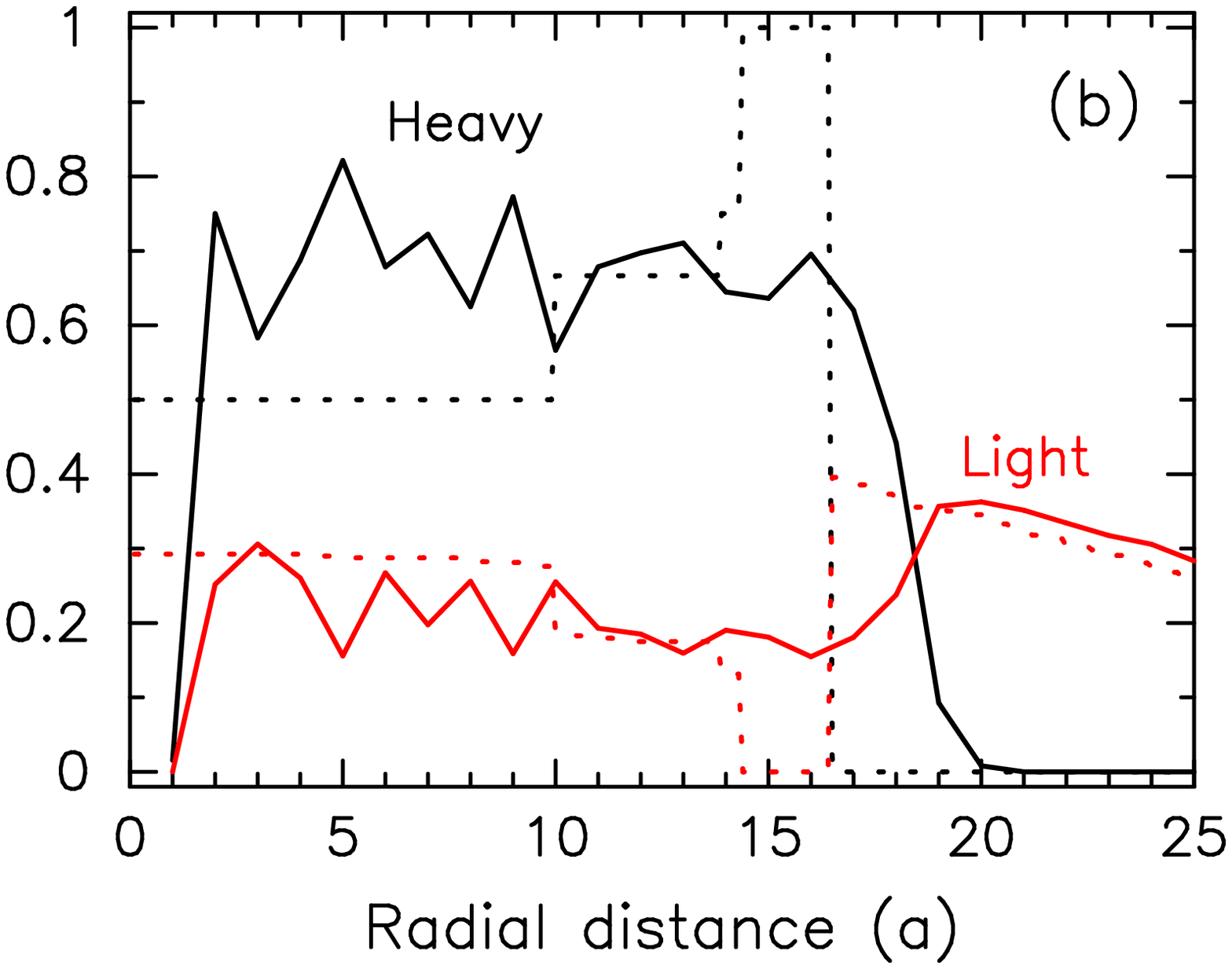}
\includegraphics[width=5.3cm]{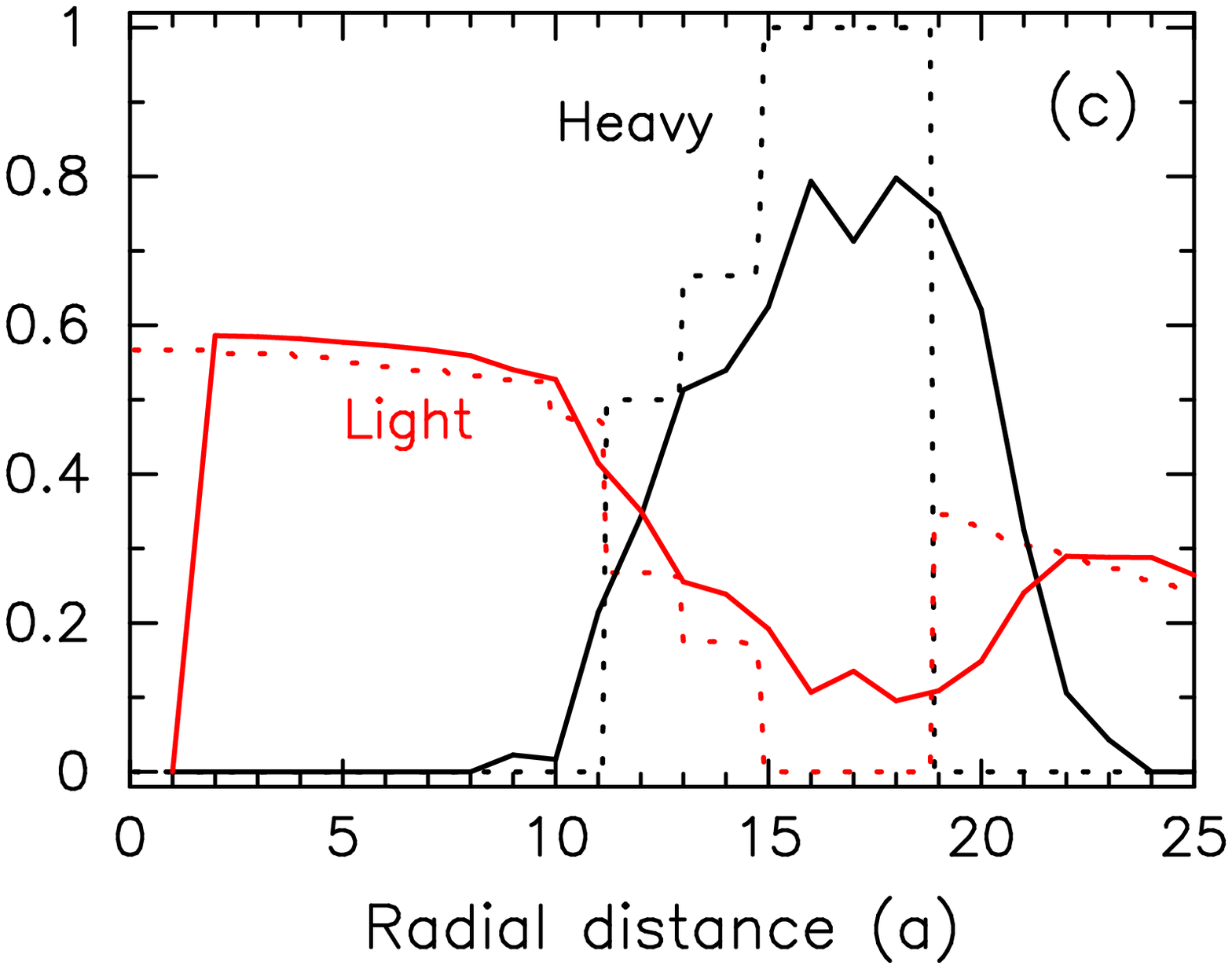}\\

\includegraphics[width=5.3cm]{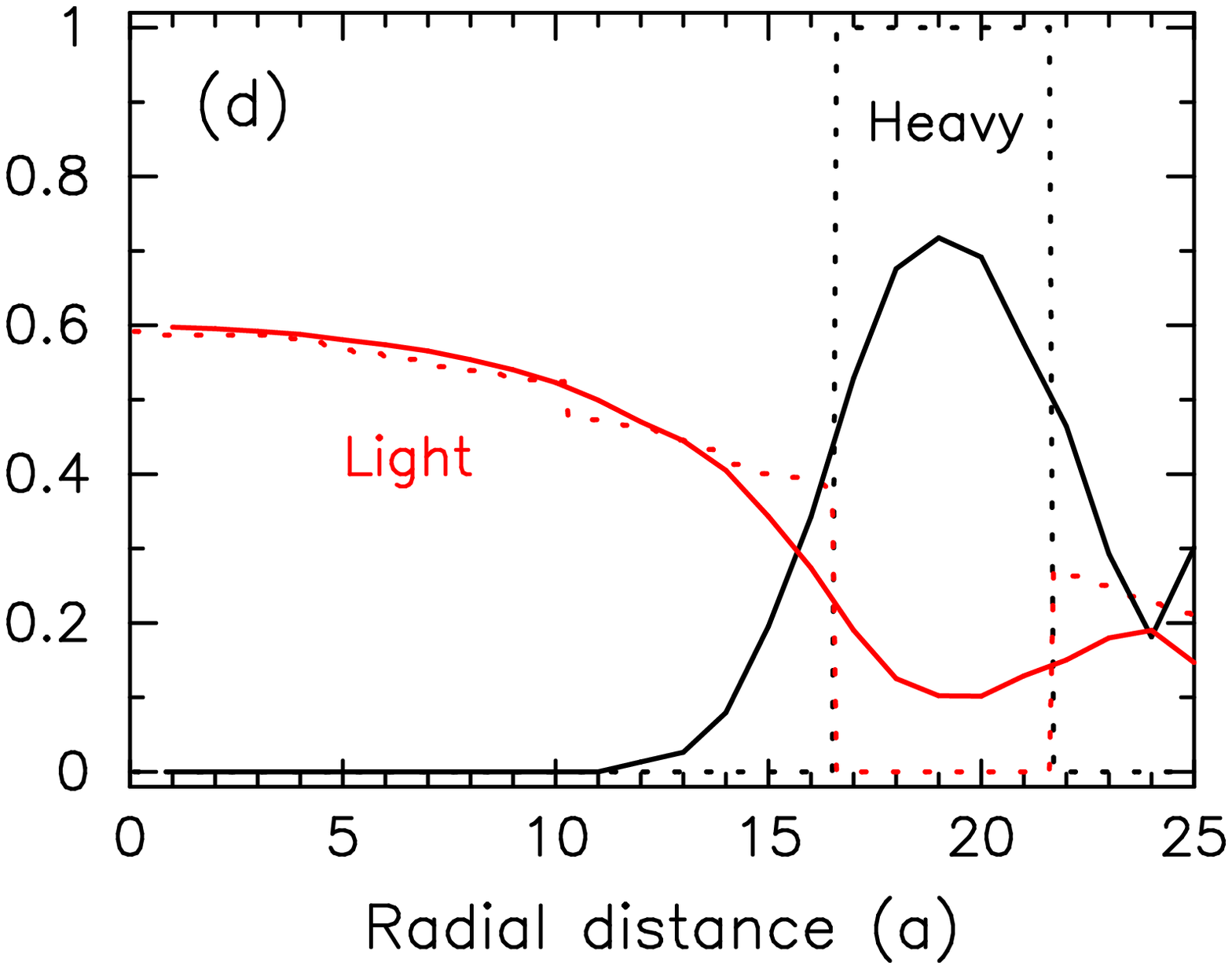}
\includegraphics[width=5.3cm]{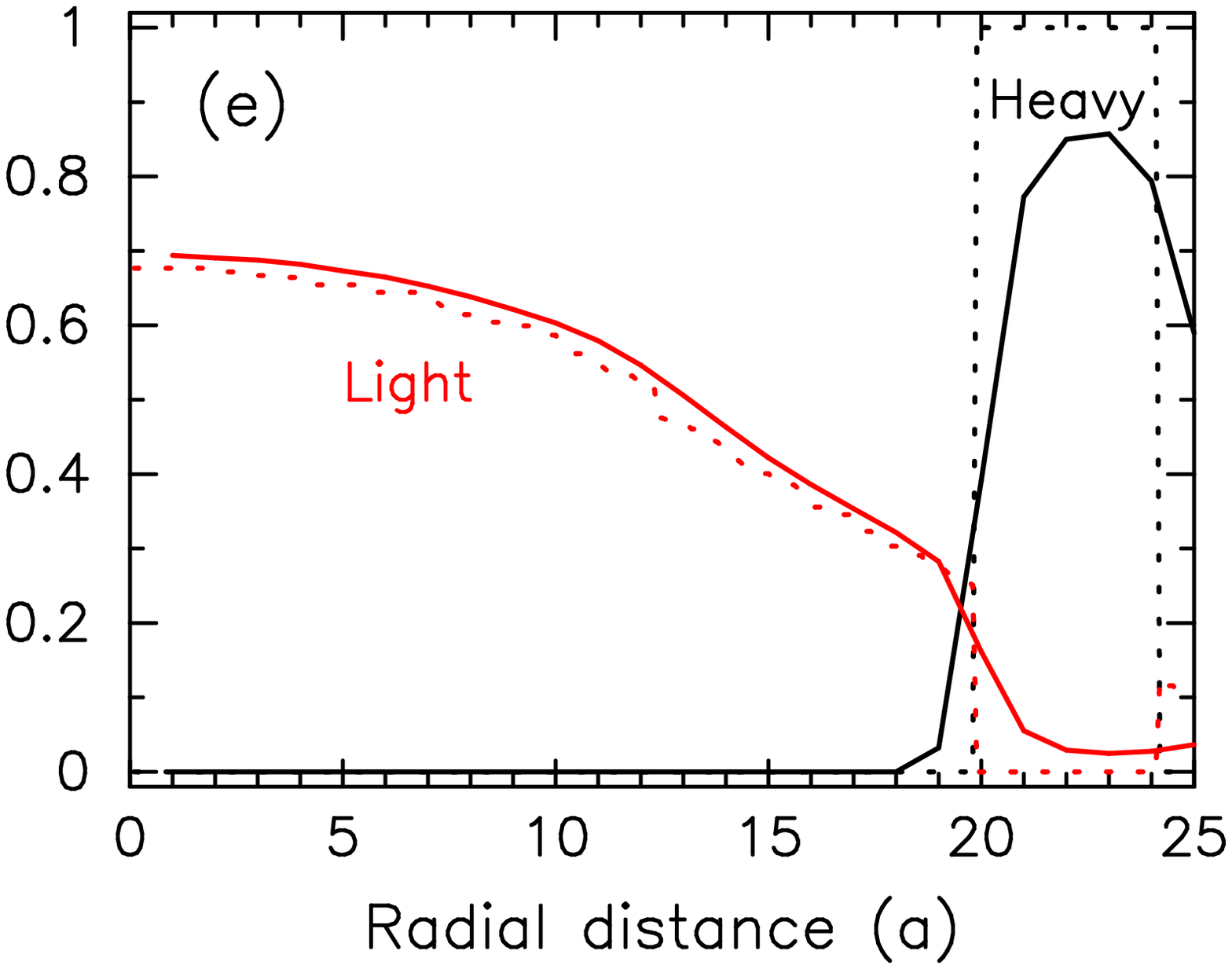}

\caption{(Color online)
Averaged heavy and light radial densities per lattice site for $U=5J$ and $k_BT=0$ in
the LDA (dotted line) compared to the QMC snapshot average at $k_BT=0.01J$ (solid line).
The five cases are identical to those shown in Fig.~\ref{fig: phase-sep}.
\label{fig: lda}}
\end{figure*}

\begin{figure*}[h]
\includegraphics[width=5.6cm]{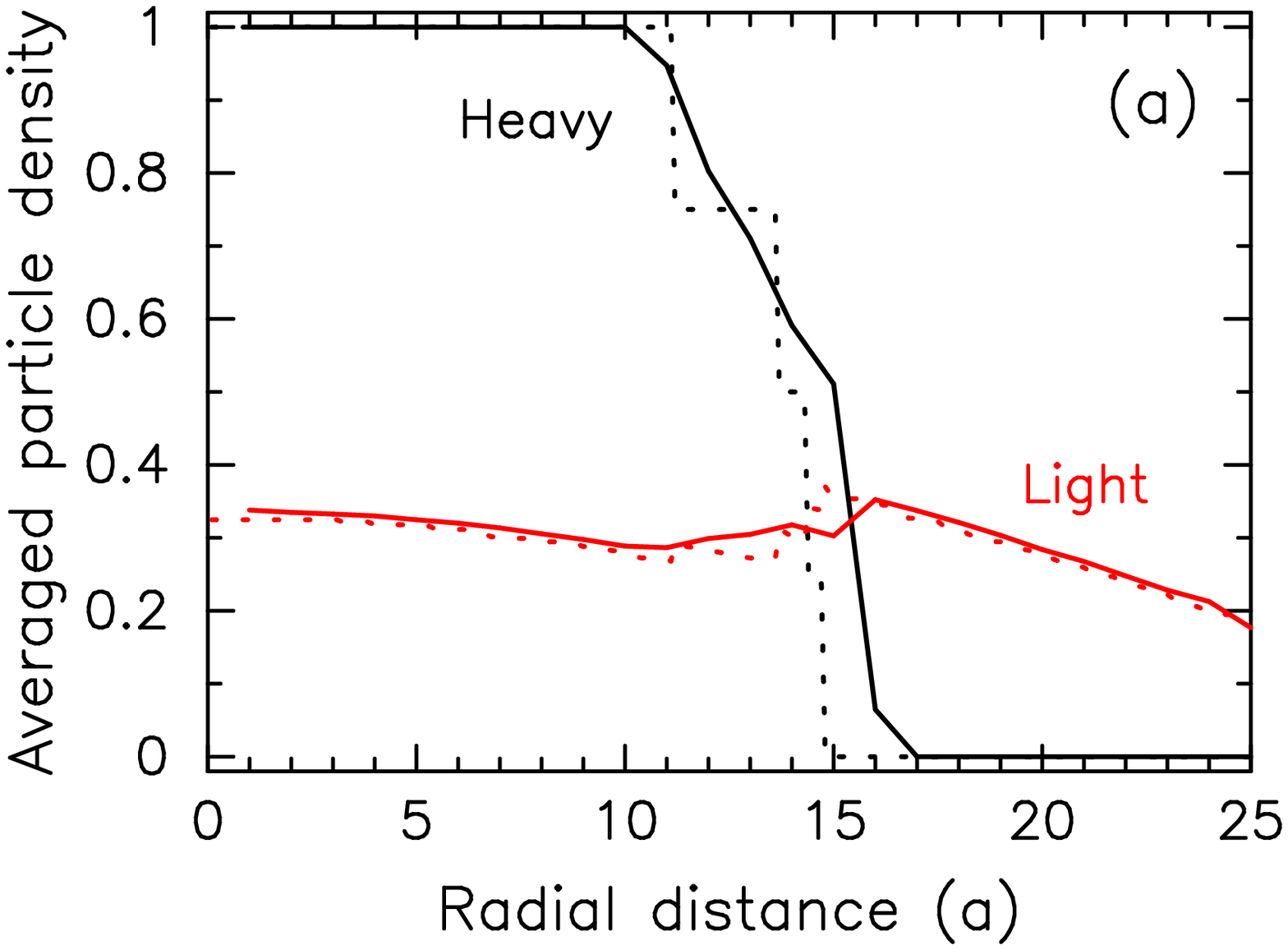}
\includegraphics[width=5.3cm]{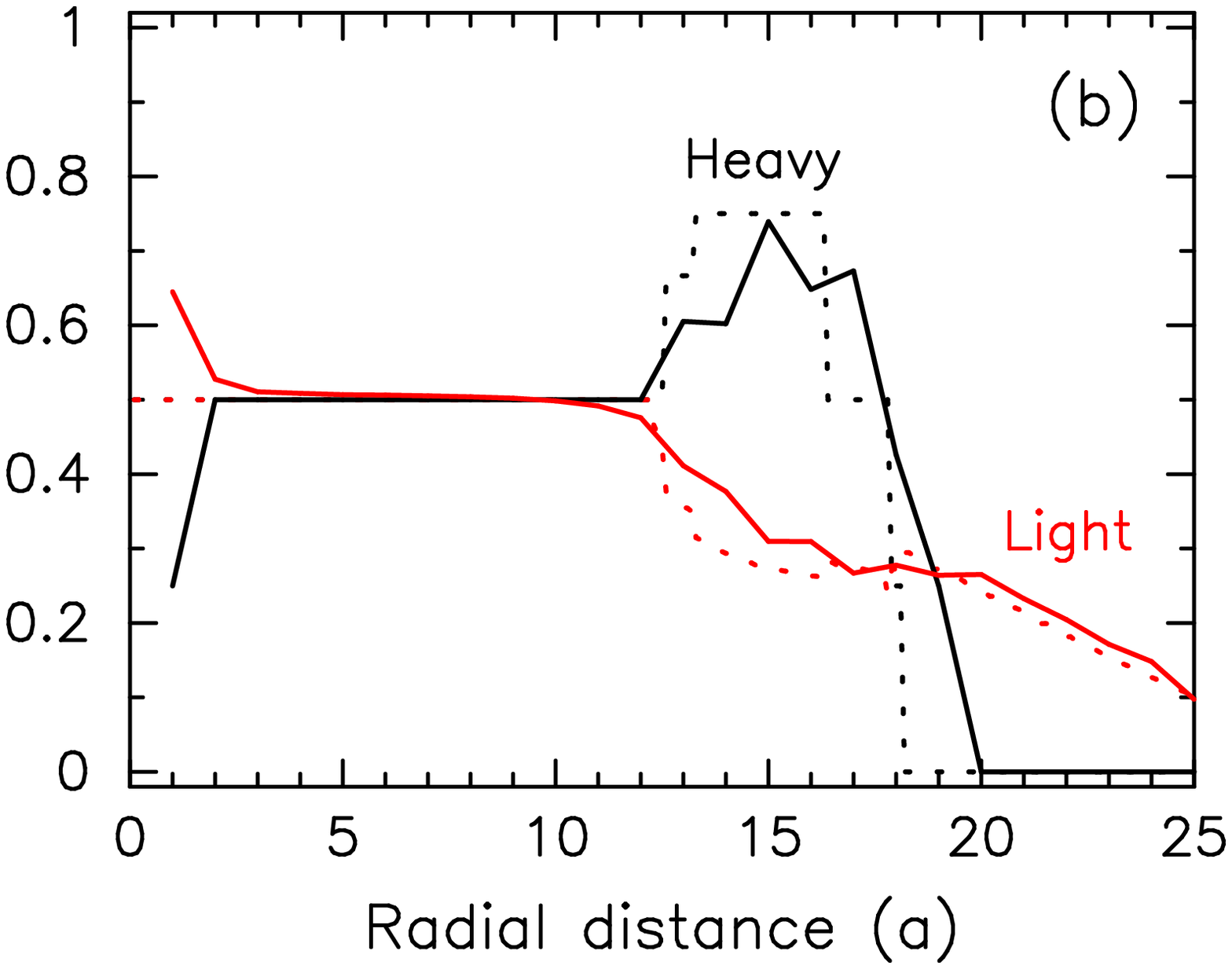}

\caption{(Color online)
Averaged heavy and light radial densities per lattice site for $U=J$ and $k_BT=0$ in
the LDA (dotted line) compared to the QMC snapshot average at $k_BT=0.0001J$ (solid line).
The two cases are (a) $\hbar\omega/2J=1/17$ and (b) $\hbar\omega/2J=1/12.9$ and correspond to the low $T$ images in Fig.~\ref{fig: u=1}.
\label{fig: lda2}}
\end{figure*}

\end{widetext}

\end{document}